\documentclass[11pt]{article}
\pdfoutput=1 
\usepackage{mathtools}
\usepackage{booktabs}
\usepackage[english]{babel}
\usepackage{amsmath,amssymb,amsbsy,amstext, amsthm, simplewick, amsfonts,braket}
\usepackage{graphicx}
\usepackage[small]{caption}
\usepackage{siunitx}
\usepackage{upgreek}
\usepackage{framed}
\usepackage{wrapfig}
\usepackage{multirow}
\usepackage{bbm}
\usepackage[numbers,sort&compress]{natbib}
\usepackage[svgnames,dvipsnames,x11names]{xcolor}
\usepackage[utf8x]{inputenc}
\usepackage{selinput}
\usepackage{bm}
\usepackage{float}
\usepackage{dsfont}
\usepackage{caption}
\usepackage{subcaption}
\usepackage{sidecap}
\usepackage{longtable}
\usepackage{anyfontsize}

\usepackage{epstopdf}
\usepackage{cancel}
\usepackage{tcolorbox}
\usepackage{latexsym,amsmath,amssymb,epsfig}
\usepackage{braket}
\usepackage{tensor}
\usepackage{tocloft}

\usepackage{tikz-cd}

\usepackage{ytableau}
\ytableausetup{centertableaux,boxsize=.8em}
\usepackage{genyoungtabtikz}
\Yboxdim{12pt} 
\Ylinethick{.9pt}

\usepackage{tcolorbox}
\definecolor{greyish2}{rgb}{.96,.96,.96}

\def\xyma{\xymatrix@M.7em}
\def\xymas{\xymatrix@M.1em}

\newcommand{\Comment}[1]{{}}
\definecolor{darkblue}{rgb}{0.15,0.35,0.55}
\definecolor{reddish}{rgb}{0.65, 0.2, 0.2}
\definecolor{darkgreen}{RGB}{50,150,0}
\definecolor{greyish2}{rgb}{.96,.96,.96}
\usepackage[linktocpage=true]{hyperref}
\hypersetup{
colorlinks=true,
citecolor=darkblue,
linkcolor=reddish,
urlcolor=darkblue,
pdfauthor={},
pdftitle={},
pdfsubject={}
}

\DeclareFontFamily{OT1}{rsfs10}{}
\DeclareFontShape{OT1}{rsfs10}{m}{n}{ <-> rsfs10 }{}
\DeclareMathAlphabet{\mathscript}{OT1}{rsfs10}{m}{n}

\def\gsim{ \lower .75ex \hbox{$\sim$} \llap{\raise .27ex \hbox{$>$}} }
\def\lsim{ \lower .75ex \hbox{$\sim$} \llap{\raise .27ex \hbox{$<$}} }
\def\be{\begin{equation}}
\def\ee{\end{equation}}
\def\bea{\begin{eqnarray}}
\def\eea{\end{eqnarray}}

\newcommand{\baaa}{\begin{eqnarray}}
\newcommand{\eaaa}{\end{eqnarray}}

\newcommand{\rd}{{\rm d}}

\newcommand{\half}{\frac{1}{2}}
\newcommand{\nn}{\nonumber}

\usepackage[letterpaper,margin=1in]{geometry}

\usepackage{tikz}
\usetikzlibrary{decorations}
\pgfdeclaredecoration{complete sines}{initial}
{
    \state{initial}[
        width=+0pt,
        next state=upsine,
        persistent precomputation={\pgfmathsetmacro\matchinglength{
            \pgfdecoratedinputsegmentlength / int(\pgfdecoratedinputsegmentlength/\pgfdecorationsegmentlength)}
            \setlength{\pgfdecorationsegmentlength}{\matchinglength pt}
        }] {}
    \state{upsine}[width=\pgfdecorationsegmentlength,next state=downsine]{
        \pgfpathsine{\pgfpoint{0.25\pgfdecorationsegmentlength}{0.5\pgfdecorationsegmentamplitude}}
        \pgfpathcosine{\pgfpoint{0.25\pgfdecorationsegmentlength}{-0.5\pgfdecorationsegmentamplitude}}
    }
    \state{downsine}[width=\pgfdecorationsegmentlength,next state=upsine]{
        \pgfpathsine{\pgfpoint{0.25\pgfdecorationsegmentlength}{-0.5\pgfdecorationsegmentamplitude}}
        \pgfpathcosine{\pgfpoint{0.25\pgfdecorationsegmentlength}{0.5\pgfdecorationsegmentamplitude}}
}
    \state{final}{}
}

\definecolor{greyish}{rgb}{.90,.90,.90}
\definecolor{greyish2}{rgb}{.96,.96,.96}
\usepackage{xcolor,colortbl}
\usepackage{tcolorbox}

\usepackage[all]{xy}


\numberwithin{equation}{section}

\begin{document}

%
%\maketitle
\renewcommand{\thefootnote}{\fnsymbol{footnote}}
%~
\vspace{0truecm}
\thispagestyle{empty}

\begin{center}
{\fontsize{20}{24} \bf A Partially Massless Superconductor}
\end{center}

%\vspace{.005truecm}

\vspace{-4pt}

\begin{center}
{\fontsize{14}{18}\selectfont
Kurt Hinterbichler${}^{\rm a,}$\footnote{\texttt{\href{mailto:kurt.hinterbichler@case.edu}{kurt.hinterbichler@case.edu}}}
and 
Austin Joyce${}^{\rm b,}$\footnote{\texttt{\href{mailto:austinjoyce@uchicago.edu}{austinjoyce@uchicago.edu}}}
}
\end{center}

\vspace{.25truecm}

 \centerline{{\it ${}^{\rm a}$CERCA, Department of Physics,}}
 \centerline{{\it Case Western Reserve University, 10900 Euclid Ave, Cleveland, OH 44106, USA}} 
 
\vspace{.5cm}

   \centerline{{\it ${}^{\rm b}$Kavli Institute for Cosmological Physics, Department of Astronomy and Astrophysics}}
 \centerline{{\it University of Chicago, Chicago, IL 60637, USA} }

 \vspace{.25cm}

\vspace{.3cm}
\begin{abstract}
\noindent 
We describe a Higgs mechanism for the partially massless graviton. In order to do so, we first construct a covariant fracton-like effective field theory on de Sitter space that linearly realizes a dipolar shift symmetry. The global symmetry of this theory can be gauged by coupling it to a partially massless spin-2 gauge field. When the fractonic matter condenses, the dipole symmetry is spontaneously broken, the resulting Goldstone mode is a galileon, and the partially massless graviton combines with this mode and becomes fully massive. At long distances, the phenomenology of this theory is captured by a quasi-topological field theory and displays many features analogous to those of the superconducting phase in electromagnetism, including gapless edge modes and persistent currents. We also describe the generalization to higher spins.

\end{abstract}

\newpage

%\pagenumbering{arabic}
\setcounter{page}{2}
\setcounter{tocdepth}{2}
\tableofcontents
\newpage
\renewcommand*{\thefootnote}{\arabic{footnote}}
\setcounter{footnote}{0}

\section{Introduction}
\label{sec:intro}

One way that we learn about theories is to deform them and study their phase transitions. The symmetries and order parameters that define different phases can often be used as a means to characterize the theory itself. This is the essence of the Landau paradigm, which aims to describe all possible phases and the transitions between them in terms of symmetries and their realization.

We have been able to understand many of the features of our world through the lens of the Landau paradigm. Indeed, essentially all of the light degrees of freedom we encounter can be understood in terms of broken symmetries.  There is, however, a notable exception---that of gravity. Despite being the most obvious force in our everyday lives, and the one that shapes the universe on the largest scales, a complete understanding of Einstein gravity in terms of global symmetries remains elusive.\footnote{See however~\cite{Benedetti:2021lxj,Hinterbichler:2022agn,Benedetti:2023ipt,Gomez-Fayren:2023qly,Hinterbichler:2024cxn,Hull:2024xgo,Hull:2024bcl,Hull:2024ism} for recent discussions of the symmetries of linearized gravity (which are also symmetries of the deep infrared of Einstein gravity \cite{Farnsworth:2021zgj}) as well as~\cite{Cheung:2024ypq} for some symmetries in fully nonlinear GR.} There are many reasons that such a description is desirable: it could provide insights into the emergence of gravity at low energies from the ultraviolet, or shed new light on the infrared structure of gravity.
Perhaps most intriguingly, a Landau theory understanding of gravity would make it possible to consider other phases of gravity besides the one that we find ourselves in. Such exotic phases of gravity could make for new and potentially interesting early universe scenarios~\cite{Agrawal:2020xek}.

Given the importance of understanding gravity and its phases, insights from any direction are valuable. In this paper, we study an analogous problem in a somewhat simpler setting: we explore the phases of the partially massless (PM) graviton~\cite{Deser:1983tm,Deser:1983mm,Higuchi:1986py}. The PM graviton is the simplest of a family of higher-spin representations in de Sitter (dS) space that do not have flat-space counterparts~\cite{Brink:2000ag,Deser:2001pe,Deser:2001us,Deser:2001wx,Deser:2001xr,Zinoviev:2001dt,Garidi:2003ys,Skvortsov:2006at,Skvortsov:2009zu}.  
It is a spin-2 field, but shares many features with electromagnetism: it enjoys a scalar gauge invariance, has a single-derivative gauge-invariant field strength \cite{Deser:2006zx}, propagates on the light cone~\cite{Deser:2001xr}, and has duality invariance in $D=4$~\cite{Deser:2013xb,Hinterbichler:2014xga}.
We know that electromagnetism has a Higgs phase inside a superconductor. There, the photon is gapped via the abelian Higgs mechanism, and the photon gets a longitudinal mode by mixing with a Goldstone mode.  This Goldstone mode is, in the global symmetry limit, a massless scalar with a shift symmetry.  Given the similarities between the PM field and electromagnetism, it is natural to ask whether it is possible to Higgs the PM field and induce a PM superconducting phase.  

Here we describe such a Higgsing of the partially massless graviton. The slightly peculiar features of the PM field relative to electromagnetism require a similarly exotic type of matter in order to gap it. The gauge invariance of the PM graviton involves two derivatives, and so the corresponding global symmetries involve a dipole-like transformation, linear in the spacetime coordinates.   We therefore require a matter sector with a spacetime dipole symmetry.  In the non-relativistic context, theories of this type are called fractons, and have been of much interest in recent years~\cite{Nandkishore:2018sel,Pretko:2020cko}. 

We first construct a relativistic version of a fracton theory---defined in de Sitter space---and then couple it to the partially massless graviton.\footnote{Similar covariant constructions in flat space can be found in \cite{Wang:2019aiq,Bertolini:2022ijb,Blasi:2022mbl,Afxonidis:2023pdq,Rovere:2024nwc,Rovere:2025nfj,Makino:2025mzo}, and in more general contexts in \cite{Jain:2021ibh,Bidussi:2021nmp,Pena-Benitez:2021ipo}. Similarities to the partially massless fields were noted in \cite{Bidussi:2021nmp}.} When the fractonic matter condenses, the dipole symmetry becomes non-linearly realized, where it is known as a galileon symmetry~\cite{Nicolis:2008in}.  The galileon Goldstone mode then combines with the partially massless field, leaving a fully massive spin-2 field and a scalar radial mode, thus producing an analogue of a superconducting phase.

This construction can be viewed as a Landau--Ginzburg description of the phase transition between a partially massless and a fully massive gravity phase.   This massive gravity theory is of the pseudo-linear type~\cite{Folkerts:2011ev,Hinterbichler:2013eza} built from linearized gravity, rather than the nonlinear dRGT type~\cite{deRham:2010kj} built from GR, but nevertheless represents an interesting proof-of-principle. 

In the Higgs phase, all degrees of freedom are massive and can be parametrically heavier than the Hubble scale. It then makes sense to integrate out all the propagating degrees of freedom in order to obtain a topological effective description of the ground states of the theory. Such a description is a BF-like theory, involving mixed symmetry gauge fields, which source the magnetic higher-form symmetry current of the partially massless field and the global symmetry current of the fractons. This theory has a number of interesting features: it supports edge modes localized on the future boundary of de Sitter space, and captures the phenomenology of persistent dipole currents in the superconducting phase.

The partially massless graviton is the first in a family of maximal depth partially massless higher-spin fields in de Sitter space. Each of these fields also possesses a scalar gauge invariance, with reducibility parameters that are higher multipoles. These fields can be Higgsed in a similar manner to the spin-2 case, albeit using higher-derivative fractonic theories.  We sketch this construction, along with the related topological field theory descriptions.

Aside from providing some insights into what the phases of a gravitational theory can look like, aspects of the construction are of independent interest.
Foremost, it is a covariant Higgsing of a non-vector theory, which is novel field-theoretically. In addition to this, the relativistic fracton theories that we consider may be of independent phenomenological interest. Further, the construction provides a UV extension of theories of massive higher spins in de Sitter space, and the properties of the topological gauge theories and their edge modes could have observational relevance.

\vspace{-8pt}
\paragraph{Outline:} An outline of this paper is the following. In Section~\ref{sec:PMandfractons} we introduce the fractonic matter theory on dS and describe its symmetry breaking phase.  In Section~\ref{sec:PMHiggs} we describe how to couple these fractons to partially massless gravity in order to gauge the symmetries. We then study how the symmetry breaking phase Higgses the graviton. We comment on the relation to massive gravity, and then describe the low-energy BF theory that describes the Higgs phase when all the degrees of freedom are integrated out. In Section~\ref{sec:HS} we describe the generalization of the construction to maximal-depth partially massless higher spins. We conclude in Section~\ref{sec:conclusions}.

\vspace{-8pt}
\paragraph{Conventions:} We work in $D$-dimensional de Sitter space, with Hubble parameter $H$ (radius $1/H$), and mostly plus metric signature. The curvatures are $R_{\mu\nu\rho\sigma} = H^2\left(g_{\mu\rho}g_{\nu\sigma} - g_{\mu\sigma}g_{\nu\rho}\right)$, 
$R_{\mu\nu} = (D-1)H^2 g_{\mu\nu}$, $R = D(D-1)H^2$.
We (anti-)symmetrize indices with weight one, so that for example $S_{(\mu\nu)} = \frac{1}{2}(S_{\mu\nu}+S_{\nu\mu})$. We define the depth, $t$, of partially massless fields to be the highest helicity removed by the gauge invariance (equivalently, the number of indices of the gauge parameter).

\section{De Sitter fractons}
\label{sec:PMandfractons}

 In order to describe the transition between the Coulomb and Higgs phases of the partially massless graviton, we will need to introduce a relatively exotic form of matter. Fortunately this class of theories---fractons---has been studied in great depth in recent years.
We will require a relativistic version of these theories. 
In this section we introduce the fracton-like matter fields that we will later couple to partially massless gravity.

\subsection{Relativistic fractons in de Sitter space}
\label{sec:fracton}

Gauge fields naturally couple to matter sectors that have as global symmetries the reducibility parameters of the gauge fields' gauge invariance. Recall that the reducibility parameters are the values of the gauge parameter for which the gauge transformation vanishes. For example, the reducibility parameters of electromagnetism are the constants which parametrize a U$(1)$, and thus electromagnetism naturally couples to matter sectors with a global U$(1)$ charge symmetry. 

As we review in Section~\ref{PMsymmsection}, the gauge transformation of a partially massless spin-2 field, $h_{\mu\nu}$, is $\delta h_{\mu\nu} = (\nabla_\mu\nabla_\nu+H^2 g_{\mu\nu})\alpha(x)$, with $\alpha(x)$ an arbitrary function. We are therefore looking for a matter sector consisting of a complex scalar field $\Phi$ that has a symmetry of the form 
\be 
\Phi \mapsto e^{i\alpha(x)}\Phi\,,\label{eq:fractonsymm}
\ee
where $\alpha(x)$ is a solution to 
\be 
\left(\nabla_\mu\nabla_\nu+H^2 g_{\mu\nu}\right)\alpha(x) = 0\,.\label{eq:fractonreducibility}
\ee
The solutions of this equation are the reducibility parameters of a PM gauge field.  There are $D+1$ independent such solutions.\footnote{We can enumerate the solutions to this equation covariantly by embedding de Sitter space as a hyperboloid in an ambient space of one higher dimension~\cite{Dirac:1936fq,Fronsdal:1978vb}. Given such an embedding $x^\mu \hookrightarrow X^A$, with $X^A$ the ambient space coordinates, the equation $\left(\nabla_\mu\nabla_\nu+H^2 g_{\mu\nu}\right)\alpha(x) = 0$ can be written as $\partial_A\partial_B \alpha(X) = 0$, which has a finite number of solutions $\alpha(X) = c_AX^A$, with $c_A$ a constant ambient space vector.}  Note also that the $\alpha(x)$ that satisfy~\eqref{eq:fractonreducibility} are precisely the shift symmetries of the galileon on dS space~\cite{Goon:2011uw,Goon:2011qf,Burrage:2011bt} (or the $k=1$ shift symmetric scalar, in the language of~\cite{Bonifacio:2018zex,Bonifacio:2021mrf}), so we can think of~\eqref{eq:fractonsymm} as a linear realization of the galileon symmetry on dS.  Theories with such linearly realized dipolar symmetries in the spatial coordinates have been widely studied in recent years in connection to condensed matter systems with fracton excitations~\cite{Nandkishore:2018sel,Pretko:2020cko}. Indeed the Higgs mechanism that we will describe can be thought of as a relativistic cousin of the constructions of~\cite{Ma_2018,Bulmash:2018lid}.

In order to implement the symmetry~\eqref{eq:fractonsymm}, we define a covariant derivative,
\be
\nabla^2_{\mu\nu}\Phi^2 \equiv \Phi^2 \left(
\nabla_\mu\nabla_\nu+H^2 g_{\mu\nu}\right)\log\left(\frac{\Phi}{f}\right)\,,
\label{eq:covariantD}
\ee
where $f$ is an arbitrary constant (which we will eventually set to a convenient value).  As we will see later, we take the field $\Phi$ to have a non-standard mass dimension $[\Phi]=D/4$, and so $f$ also has mass dimension $[f]=D/4$.
In the limit $H\to 0$, this covariant derivative reduces to (a relativistic version of) that of~\cite{Pretko:2018jbi}.  In the flat limit the logarithm disappears after taking the derivatives, but in dS space the log persists.
Under the transformation \eqref{eq:fractonsymm}, with an arbitrary function $\alpha(x)$, the covariant derivative shifts as $\nabla^2_{\mu\nu}\Phi^2  \mapsto e^{2i\alpha(x)} \nabla^2_{\mu\nu}\Phi^2 +ie^{2i\alpha(x)}\Phi^2\left( \nabla_\mu\nabla_\nu+H^2 g_{\mu\nu}\right)\alpha(x).$
From this, we see that if the function $\alpha(x)$ satisfies \eqref{eq:fractonreducibility}, then $\nabla^2_{\mu\nu}\Phi^2$ transforms covariantly by a phase,
\be
\nabla^2_{\mu\nu}\Phi^2  \mapsto e^{2i\alpha(x)} \nabla^2_{\mu\nu}\Phi^2 \,.
\ee
 Given this covariant derivative, it is then straightforward to construct invariant actions by combining with its complex conjugate and contracting indices in all possible ways.\footnote{In addition to manifestly invariant terms built from $\nabla^2_{\mu\nu}\Phi^2$, there is also a Wess--Zumino term, which shifts by a total derivative under the transformation~\eqref{eq:fractonsymm}:
\be
S_{\rm wz} \propto \int\rd^Dx \sqrt{-g}\Bigg[\log\left(\frac{\Phi}{f}\right)\big(\square+DH^2\big) \log\left(\frac{\Phi}{f}\right)+{
\rm c.c.}
\Bigg]\,.
\label{eq:fractonicwzterm}
\ee
In order for this term to be invariant, we actually only require that the trace of~\eqref{eq:fractonreducibility} be satisfied, $(\square+DH^2)\alpha(x) = 0$. As such, this Wess--Zumino term has more symmetry than the terms constructed from~\eqref{eq:fractonreducibility}. Since we will ultimately be interested in gauging the full set of symmetries by coupling to a partially massless field, we do not include this Wess--Zumino term in the following.}

As a consequence of the dipolar symmetry~\eqref{eq:fractonsymm}, these fractonic theories possess a two-index Noether current $J_{\mu\nu}$ that satisfies the conservation condition $\left(\nabla_\mu\nabla_\nu+H^2g_{\mu\nu}\right) J^{\mu\nu} = 0$ on shell.  This current will couple to the partially massless gauge field once the fractonic symmetry is gauged.

\subsection{Symmetry breaking\label{symbreakingsubsection}}

As in the ordinary Higgs mechanism, we want the matter Lagrangian to have a symmetry breaking phase in order to be able to gap the gauge sector. This can be accomplished with the following action
\be
S = 
\int\rd^Dx \sqrt{-g}\Bigg[ -\frac{1}{M^4}\Big(\lvert \nabla^2_{\mu\nu}\Phi^2\rvert^2-\lvert \nabla^2{}_\mu^{~\mu}\Phi^2\rvert^2\Big)+\frac{1}{M^2}\left(
\Phi^*{}^2\nabla^2{}_\mu^{~\mu}\Phi^2+{\rm c.c.}\right)+\mu^{D/2}\lvert\Phi\rvert^2-\frac{\lambda}{2}\lvert\Phi\rvert^4
\Bigg]\,.
\label{eq:fractonaction}
\ee
This action has a number of features that deserve explanation. First, there is no standard quadratic kinetic term for $\Phi$; such a term is not allowed by the fractonic symmetry.  The absence of a standard kinetic term, as well as the presence of the log in the covariant derivative \eqref{eq:covariantD}, means that the theory is not well defined around $\Phi = 0$.  It nevertheless can be expanded around nontrivial backgrounds to describe small fluctuations around these states.\footnote{In this respect the theory is similar to effective field theories of condensed matter systems like fluids~\cite{Dubovsky:2011sj,Liu:2018kfw}, fermions at unitarity~\cite{Son:2005rv}, large charge CFT \cite{Hellerman:2015nra}, or other exotic systems like the ghost condensate~\cite{Arkani-Hamed:2003pdi}.} 
Secondly, there are two mass scales $M$, $\mu$, the scale $f$ in the log, and a dimensionless coupling $\lambda$, all of which we will take to be positive.  The mass dimension of the field is $[\Phi] = \tfrac{D}{4}$.  The scales are organized so that $M$ suppresses powers of derivatives, and $\mu$ comes with powers of the field.  We can imagine that a Lagrangian like this comes from integrating out more fundamental degrees of freedom at the scale $M$, and so we consider this to be a Landau--Ginzburg-like effective field theory (EFT) with a cutoff scale $M$, where we have kept interactions up to four derivative order.  In order for the background de Sitter scale to sit below this cutoff, we will require
\be {H\over M}\ll 1.\label{HsmallerthanMcufee}\ee
Given that this is an EFT, and that EFTs should have all possible terms compatible with the symmetries, there seem to be several possible dimensionless coefficients that are missing in~\eqref{eq:fractonaction}.  Some are due to redundancies: among the terms quartic in the fields, the coefficients in front of the two-derivative part can be adjusted by changing the scale $f$ of the log and redefining $\lambda$.  This has been used to remove any dimensionless coefficient in front of the four-derivative terms relative to the two-derivative terms.  The scale $f$ of the log can then be set to any value by rescaling the field and redefining the scales $M$, $\mu$, and the coupling $\lambda$.  We will use this freedom later to set $f$ to a convenient value.  There are two genuine tunings not due to redundancies that we have made: the coefficient in front of the two-derivative term could be complex, but if it had an imaginary part, we would not find the symmetry-breaking solution discussed below, so we have tuned the imaginary part to zero.  Additionally, the four-derivative terms in~\eqref{eq:fractonaction} have a relative Fierz--Pauli-like tuning.  This ensures that the equations of motion of the theory are second order (so that there are no extra propagating ghostly modes), which will in turn guarantee that the Goldstone field in the symmetry-breaking phase has an ordinary two-derivative kinetic term, and that the massive graviton that emerges once we couple to the PM field has a Fierz--Pauli mass term.

We now search for a constant solution to the equations of motion following from~\eqref{eq:fractonaction}. Taking $\Phi = v$, with $v>0$ and constant, the equations of motion reduce to
\be
M^{\frac{D-2}{2}}\mu^2 v + (DH^2-M^2\lambda)v^3+\frac{DH^2}{M^2}\log\left(\frac{v}{f}\right) v^3\left[4M^2+(D-1)H^2+2H^2(D-1)\log\left(\frac{v}{f}\right)\right]=0\,.
\ee
Solving this equation exactly is rather difficult, but we can solve it perturbatively as an expansion in powers of $H$.  At this point, we use the freedom mentioned above to choose the scale appearing in the logs to be 
\be f = \mu^\frac{D}{4} /\sqrt\lambda.\ee
Making this choice eliminates logs in the expansion of the solution, and the result can be written as a double power series in two dimensionless parameters, ${H^2\over M^2\lambda}$ and $\lambda$,
\be
v  = \frac{\mu^\frac{D}{4}}{\sqrt\lambda}+{\cal O}\left({H^2\over M^2\lambda},\lambda\right)\,.
\label{eq:VEV}
\ee
So that we can use only the leading order result, we will require the hierarchy 
\be {H^2\over M^2} \ll \lambda \ll 1,\label{mlbhmlmeiree}\ee
which is consistent with \eqref{HsmallerthanMcufee}.
Taking this small value of $\lambda$ is a fine tuning, which will show up later as an unnaturally small value of the mass of one of the scalars in the theory (the radial mode) compared to the cutoff $M$.

Expanding about this symmetry-breaking solution,
\be
\Phi = (v+\sigma)e^{i\phi}\,,
\label{eq:symmexpan}
\ee
we find the quadratic action for the real-valued fluctuation fields $\sigma$ and $\phi$,
\be
S = \int\rd^Dx\sqrt{-g}\left[
-\frac{1}{2}Z^2_\sigma\Big( (\nabla\sigma)^2+m_\sigma^2\sigma^2
\Big)
-\frac{1}{2}Z^2_\phi\Big( (\nabla\phi)^2-DH^2\phi^2
\Big)
\right]\,.
\ee
Here the kinetic normalizations are (where the ellipses denote terms suppressed by further powers of either ${M^2\over H^2\lambda}$ or $\lambda$),
\begin{align}
\label{eq:znorm}
Z^2_\sigma & = \frac{16\mu^{D/2}}{M^2\lambda}+\cdots \,,\\
Z^2_\phi & = 2(D-1)\frac{\mu^DH^2}{M^4\lambda^2}+\cdots\,,
\end{align}
and both kinetic terms are positive.
The mass of the $\sigma$ field is
\be
m^2_\sigma = \frac{\lambda M^2}{4}+\cdots\,,
\label{eq:radialmass}
\ee
and the mass of the $\phi$ field is exactly
\be
m^2_\phi =-DH^2\,.
\label{Goldstonemasse}
\ee

We identify $\phi$ as the Goldstone mode and $\sigma$ as the radial mode. We can see this by noting that the transformation \eqref{eq:fractonsymm}, which infinitesimally reads
\be 
\delta \Phi=i\alpha(x)\Phi,
\ee
acts on the fluctuations as
\be \delta \sigma=0\,,\ \ \quad\ \delta \phi= \alpha(x)\,,\label{Goldstoneshiftse}\ee
so that $\phi$ shifts, nonlinearly realizing the fractonic symmetry of the action~\eqref{eq:fractonaction}.  This shift is precisely the dS galileon shift symmetry of~\cite{Goon:2011uw,Goon:2011qf,Burrage:2011bt} (or the $k=1$ scalar shift symmetry of \cite{Bonifacio:2018zex,Bonifacio:2021mrf}).

Note that the Goldstone mode has a mass, which is fixed to~\eqref{Goldstonemasse} by the shift symmetry~\eqref{Goldstoneshiftse}. 
This mass is tachyonic, but as discussed in~\cite{Bros:2010wa,Epstein:2014jaa,Bonifacio:2018zex}, a field with this mass carries a 
unitary representation of the dS group.
In contrast, the mass squared of the radial mode is positive and, due to \eqref{mlbhmlmeiree}, parametrically larger than that of the Goldstone mode, as is the case in the usual Higgs mechanism,
\be m^2_\sigma \sim \lambda M^2 \gg H^2\sim m_\phi^2\,.\label{separationoffbgeee}\ee
At the same time, this mass is parametrically smaller than the original cutoff $M$,
\be m^2_\sigma \sim \lambda M^2\ll M^2.\ee
 The small Goldstone mass is natural because of the shift symmetry~\eqref{Goldstoneshiftse}, but the small mass of the radial mode relative to the cutoff is not, and is a result of the tuning $\lambda\ll1$.  In the usual Higgs mechanism, there is no explicit cutoff, but implicitly the same issue is present, as it is whenever there are light scalars unprotected by a symmetry, since one must tune the parameters to be near a phase transition.

\paragraph{Interactions:} Expanding to higher order in powers of the fields, we can find the interactions of the Goldstone mode $\phi$ and the radial mode $\sigma$. First we canonically normalize the fields by defining 
\be \hat\phi \equiv Z_\phi \phi, \ \ \quad \ \hat \sigma \equiv  Z_\sigma \sigma.\ee  

 As a consequence of the shift symmetry \eqref{Goldstoneshiftse} of the Goldstone field, interactions involving $\phi$ organize themselves in terms of the combination
\be\label{shiftinfielstreeee}
\nabla^2_{\mu\nu}\phi \equiv (\nabla_\mu\nabla_\nu+H^2 g_{\mu\nu})\phi\,.
\ee
This is a ``field strength" for the Goldstone \cite{Bonifacio:2018zex}; it is invariant under the shift symmetry \eqref{Goldstoneshiftse}.   Note also that there will be no terms odd in $\phi$, due to the $\phi\rightarrow -\phi$ symmetry stemming from charge conjugation symmetry of the original field $\Phi$.

In terms of the combination \eqref{shiftinfielstreeee}, and to leading order in $H^2/M^2$, the interactions are
\begin{align}
\nonumber
S_{\rm int} = \!\int\!\rd^Dx\sqrt{-g}\Bigg[&\frac{\lambda M H^{-2}}{2(D-1)\mu^{D/2}}\!\left(\hat\sigma+  \frac{3M\lambda}{8 \mu^{D/2}}  \hat\sigma^2 +  \frac{M^2\lambda^2}{16 \mu^{D}}  \hat\sigma^3 +  {M^3\lambda^3\over 256\mu^{3D/2}}  \hat\sigma^4   \right)\!\!\left[ (\nabla^2{}_{\mu}^\mu\hat\phi)^2-(\nabla^2_{\mu\nu}\hat\phi)^2\right]\\[2pt]\nonumber
&-\frac{\lambda^2M^3}{32\mu^{D/2}}\hat\sigma^3
-\frac{\lambda M}{4\mu^{D/2}}\hat\sigma(\nabla\hat\sigma)^2-\frac{3\lambda}{32M\mu^{D/2}}(\nabla\hat\sigma)^2\square\hat\sigma-\frac{\lambda^3M^4}{512\mu^D}\hat\sigma^4
\\[2pt]
&\!\!-\frac{\lambda^2M^2}{32\mu^D}\hat\sigma^2(\nabla\hat\sigma)^2+\frac{\lambda^2}{64\mu^D}\hat\sigma\Big(\nabla_\mu\nabla_\nu\hat\sigma\nabla^\mu\hat\sigma\nabla^\nu\hat\sigma-(\nabla\hat\sigma)^2\square\hat\sigma
\Big)+\cdots\!\Bigg]\,.
\label{eq:fractonssbinteractions}
\end{align}
The characteristic size of the action is the energy density of the background $\Phi = v$, which is
\be \rho \sim \mu^D/\lambda\,.\label{backgroundscalelee}\ee
 On top of this, we see that each power of $\hat \sigma$ comes with a factor of $\mu^{D/2}/(M\lambda)$, and each power of $\nabla^2\hat\phi$ comes with a factor of $H \mu^{D/2}/\sqrt{\lambda}$. Additional derivatives are suppressed by either $1/M$ or $1/(\sqrt \lambda M)$.  The schematic structure of the Lagrangian is thus\footnote{Note that the kinetic term for $\hat\phi$ in this structure comes from the term of the form $\sim (\nabla^2{}_{\mu}^\mu\hat\phi)^2-(\nabla^2_{\mu\nu}\hat\phi)^2$, which reduces to the standard two-derivative kinetic term, up to a total derivative.}
 \be 
 {\cal L} \sim {\mu^D\over \lambda} {\cal F}\left(  {\hat\sigma\over \mu^{D/2}/(M\lambda)},{ \nabla^2\hat\phi \over H \mu^{D/2}/\sqrt{\lambda}}, {\nabla\over M},{\nabla\over M\sqrt\lambda}   \right) ,
 \ee
where we have dropped terms that are suppressed by additional powers of $H/M$ compared to this power counting.
It is natural to require $v\ll M^{D/4}$, so that the energy scale of the field's expectation value is within the regime of validity of the theory.  This gives the following constraint on the parameters,
\be {\mu\over M}\ll \lambda^{2/D}\, . \label{vevbelowcutoffeee}\ee
It is also natural to require that the energy scale of the background energy density \eqref{backgroundscalelee} is below the cutoff, $\rho^{1/D}\ll M$, which implies $ \mu/M\ll \lambda^{1/D}$.  This is automatically satisfied given \eqref{vevbelowcutoffeee} and the fact that we are already requiring $\lambda\ll 1$.

\paragraph{Goldstone EFT:} 
Due to the separation of scales \eqref{separationoffbgeee}, we can integrate out the radial mode, after which we will be left with an EFT of the Goldstone mode $\phi$.   This Goldstone EFT will have the schematic form
\be {\cal L} \sim {\mu^D\over \lambda} {\cal F}\left(  { \nabla^2\hat\phi \over H \mu^{D/2}/\sqrt{\lambda}}, {\nabla\over M},{\nabla\over M\sqrt\lambda}   \right) .\ee
This now has the structure of a galileon-like effective theory, built from the square of the shift invariant field strength \eqref{shiftinfielstreeee}, along with higher-derivative corrections suppressed by $1/(\sqrt \lambda M)$ (appropriate after integrating out the radial mode with mass $m_\sigma \sim \sqrt \lambda M$), and $1/M$ (coming from the cutoff suppression of higher-dimension operators in the original theory).  This is an example of a galileon superfluid of the type described in \cite{Hinterbichler:2022agn,Hinterbichler:2024cxn}, here realized on dS.\footnote{See~\cite{Yuan:2019geh,Wang:2019mtt,Chen:2020jew,Gromov:2020yoc,Grosvenor:2021rrt,Fliss:2021ekk,Glodkowski:2022xje,Qi:2022seq,Afxonidis:2023tup,Glorioso:2023chm,Armas:2023ouk,Jain:2023nbf,Stahl:2023prt,Wang:2024cqn} for studies of non-relativistic fracton superfluids, symmetry breaking, and hydrodynamics.}

Looking at the leading interaction $\sim \frac{\lambda}{H^4 \mu^D} \nabla^8\hat\phi^4$, we can read off the strong coupling scale $\Lambda_s$
\be \label{scsclaee}
\Lambda_s = \left(
\frac{H^4\mu^D}{\lambda}\right)^\frac{1}{D+4}\,.
\ee
For this to be the lowest scale among all the other terms involving higher powers of the field strength and no extra derivatives, we must also require\footnote{If instead we set $\mu/H\ll \lambda^{1/D}$, then interactions with additional powers of $\nabla^2\phi$ will have lower and lower strong coupling scales, asymptoting to $\Lambda_s = H \mu^{D/2}/\sqrt{\lambda}$.
}
\be 
{\mu\over H}\gg \lambda^{1/D}\,.\label{cutoffree}
\ee
Given this, $\Lambda_s$ is the strong coupling scale of the EFT. That is, given only knowledge of the terms involving $\phi$, this is the maximum scale before which new physics would have to come in.
Note that the strong coupling scale vanishes in the flat limit: $\Lambda_s\rightarrow 0$ as $H\rightarrow 0$.   The superfluid EFT picture breaks down in this limit, so that the fracton theory is better behaved in dS than in flat space.

\paragraph{Energy Scales:} Consider now the hierarchies among the various energy scales entering the problem.  In order for the full theory to be a weakly coupled UV extension of the Goldstone EFT, we want the radial mode $\sigma$ to enter before the strong coupling scale of the Goldstone EFT is reached,
\be m_\sigma \ll \Lambda_s.\ee
In addition, we want the cutoff $M$ of the full theory to be higher than the strong coupling scale, so that the regime of validity of the full theory is larger than that of the Goldstone EFT, 
\be \Lambda_s \ll M.\ee
Taking these together and using~\eqref{scsclaee} and~\eqref{eq:radialmass}, we find the condition
\be \lambda^{{D+6}\over 2} \ll \left(\mu\over M\right)^{D}\left( H\over M\right)^4 \ll  \lambda^{}\, .\label{cutoffcondiee}\ee 

Collecting together all of the various conditions we have imposed,~\eqref{mlbhmlmeiree},~\eqref{vevbelowcutoffeee},~\eqref{cutoffree}, and~\eqref{cutoffcondiee}, we have
\be 
{H^2\over M^2} \ll \lambda \, ,\ \quad \  {\mu\over M}\ll \lambda^{2/D}\,,\ \quad \  {\mu\over M}{M\over H} \gg \lambda^{1/D}\, ,\ \quad \  \lambda^{{D+6}\over 2} \ll \left(\mu\over M\right)^{D}\left( H\over M\right)^4 \ll  \lambda \,,\label{totalsinqee}
\ee
along with $ \lambda \ll 1$.  Note that the last of \eqref{totalsinqee} is not satisfied in the flat limit where $H\rightarrow 0$ with everything else fixed. 

The constraints \eqref{totalsinqee} can all be simultaneously parametrically satisfied.  For example, taking the scalings $H/ M\sim \lambda^{t_1}$, $\mu/M\sim \lambda^{t_2}$, the requirements \eqref{totalsinqee} become
\be 2t_1>1,\qquad \ t_2>{2\over D},\qquad \ t_2-t_1<{1\over D},\qquad \ {{D+6}\over 2}> 4t_1+Dt_2>1. \label{simutineeqee}\ee 
 For any $D\geq 3$, there is a small compact region in the plane of positive $t_1,t_2$ where these constraints are all satisfied. 
 Given these parameters, the fractonic theory of $\Phi$ can be considered a UV extension of the dS galileon superfluid EFT of $\phi$, analogous to the way in which a complex scalar with a symmetry-breaking potential is a UV theory extending the $P(X)$ EFT of a superfluid.  In this case, we expect that the interactions with the radial mode $\sigma$ should raise the strong coupling scale from $\Lambda_s$ to $M$.  Note that on flat space, $S$-matrix analyticity constraints forbid a normal weakly coupled UV completion of the galileon~\cite{Adams:2006sv,Bellazzini:2020cot,Caron-Huot:2020cmc,Tolley:2020gtv}, and it is thought that something non-standard is needed to complete it~\cite{Keltner:2015xda}.   We evade this both by being on dS space (which effectively gives the galileon a mass) with interactions that grow infinitely strong if we try to take the flat limit, and by having only a UV extension rather than a full completion.

\section{Higgsing the partially massless graviton}
\label{sec:PMHiggs}

We now turn to gauging the fractonic symmetry.  The gauge field for this symmetry is a partially massless graviton, so we start by reviewing the salient features of this field.
We then show that by coupling a fractonic matter sector  to a partially massless graviton, the symmetry breaking phase causes the graviton to be Higgsed and become massive by combining with the Goldstone mode~$\phi$.

\subsection{Partially massless spin-2\label{PMsymmsection}}

The representation theory of the de Sitter group is rather intricate, and there exist families of representations that have no direct flat space counterpart~\cite{thomasds,thomasds2,Dixmier,Hirai,Hirai2,Takahashi} (see~\cite{Basile:2016aen,Sun:2021thf} for recent reviews of dS representation theory). Perhaps most intriguingly, there are spinning particles that have a number of degrees of freedom intermediate between those of massive and massless fields.  Such representations are called  partially massless~\cite{Deser:1983tm,Deser:1983mm,Higuchi:1986py,Brink:2000ag,Deser:2001pe,Deser:2001us,Deser:2001wx,Deser:2001xr,Zinoviev:2001dt,Garidi:2003ys,Skvortsov:2006at,Skvortsov:2009zu}. They can be realized by tuning the mass values of massive spinning particles to special values relative to the Hubble scale, where the theory develops gauge invariances that remove some of the propagating polarizations.  These representations provide an interesting arena where features of massless theories can be explored with fewer complications.

The simplest example of a partially massless representation is the partially massless spin-2.  A free massive spin-2 particle on dS is described by the action
 \be
 \begin{aligned}
S=\int {\rm d}^Dx\sqrt{-g}\bigg[ -{1\over 2}\nabla_\lambda h_{\mu\nu} \nabla^\lambda h^{\mu\nu}&+\nabla_\lambda h_{\mu\nu} \nabla^\nu h^{\mu\lambda}-\nabla_\mu h\nabla_\nu h^{\mu\nu}+\half \nabla_\mu h\nabla^\mu h \\
&+\left(D-1\right)H^2\left( h^{\mu\nu}h_{\mu\nu}-\half h^2\right)-\frac{1}{2}m^2(h_{\mu\nu}h^{\mu\nu}-h^2)\bigg],
\label{eq:curvedmassivelin}
\end{aligned}
\ee
where $h_{\mu\nu}$ is a symmetric tensor and $h\equiv h^\mu_{\ \mu}$. There are two distinguished values of the mass parameter $m^2$. The first is $m^2 =0$, corresponding to the linearized graviton. As is well known, at this point the theory is invariant under linearized diffeomorphisms, which removes all the polarizations except for the helicity-2 polarization. If instead we dial the mass parameter $m^2$ to\footnote{This is the lightest that a massive spin-2 field can be on de Sitter space and still be unitary, apart from the massless value. For this reason, this mass value is often called the Higuchi bound~\cite{Higuchi:1986py}. For a spin-$s$ field there are $s$ distinguished PM mass values, $s-1$ of which are below the Higuchi bound (including $m^2=0$), and one at the Higuchi bound, which is spin dependent.}
\be 
m^2=\left(D-2\right)H^2\, ,\label{masstuning}
\ee
then the theory has a scalar gauge invariance
\be 
\delta h_{\mu\nu}= \left(\nabla_\mu\nabla_\nu+H^2 g_{\mu\nu}\right)\alpha(x)\,,
\label{eq:gaugesym}
\ee
with $\alpha(x)$ an arbitrary scalar function. This gauge redundancy removes only the helicity-$0$ polarization so that the theory propagates $(D+1)(D-2)/2-1$ degrees of freedom.

As is suggested by the scalar gauge invariance~\eqref{eq:gaugesym}, the theory of a partially massless graviton shares many physical features with electromagnetism. The theory possesses a gauge-invariant field strength~\cite{Deser:2006zx},
\be 
F_{\mu\nu\rho}=\nabla_\mu h_{\nu\rho}-\nabla_\nu h_{\mu\rho},
\label{eq:PMmaxwell}
\ee
which is analogous to the Maxwell tensor of electromagnetism.
This tensor is invariant under the gauge transformation~\eqref{eq:gaugesym} and has the symmetries of the hook-shaped Young diagram $\Yboxdim{7pt}\raisebox{1.25pt}{\gyoung(~;~,~)}$, i.e. it is antisymmetric in its first two indices and vanishes when antisymmetrized in all its indices. The action~\eqref{eq:curvedmassivelin} can be written elegantly in terms of this field strength tensor as~\cite{Deser:2006zx}
\be 
S=-\frac{1}{4} \int{\rm d}^Dx\sqrt{-g}\left[F^{\lambda\mu\nu}F_{\lambda\mu\nu}-2F^{\lambda\mu}_{\ \ \ \mu} F_{\lambda\nu}^{\ \ \nu}\right]. \label{PMkineticterme}
\ee
This formulation in terms of $F_{\mu\nu\rho}$ makes the gauge invariance of the action manifest. The relative tuning between the two terms, analogous to the Fierz--Pauli tuning~\cite{Fierz:1939ix} in the mass term for generic masses in \eqref{eq:curvedmassivelin}, ensures that only a PM mode propagates \cite{Farnsworth:2024iwc}. 

\paragraph{Global symmetries:} As a consequence of the equations of motion, the field strength $F_{\mu\nu\rho}$ is  conserved in all its indices and traceless on shell
\be
\nabla^\mu F_{\mu\nu\rho} = 0\,,\qquad \  \nabla^\rho F_{\mu\nu\rho} = 0,\qquad \ F^{\nu}_{\ \mu\nu}=0\,.\label{onshellFee}
\ee
Additionally, it obeys the Bianchi identity\footnote{This can be understood as a consequence of a generalization of the de Rham complex to de Sitter space~\cite{Hinterbichler:2014xga,Hinterbichler:2016fgl}. Indeed, much as Maxwell's equations can be formulated directly in terms of $F_{\mu\nu}$, the on-shell equations of motion satisfied by a partially massless spin-2 can be formulated directly in terms of $F_{\mu\nu\rho}$~\cite{Hinterbichler:2014xga}.}
\be
\nabla_{[\lambda}F_{\mu\nu]\rho} = 0\,.\label{Fbianchie}
\ee

Given the conservation equations \eqref{onshellFee}, we can construct a conserved 2-form current by contracting with the derivative of a scalar $\alpha(x)$ that satisfies the reducibility condition \eqref{eq:fractonreducibility} \cite{Hinterbichler:2015nua},
\be
J_{\mu\nu} = F_{\mu\nu\lambda}\nabla^\lambda\alpha\,.
\ee
This is conserved on shell by virtue of the conservation conditions \eqref{onshellFee} and the reducibility condition \eqref{eq:fractonreducibility}.
Using this conserved 2-form, we can construct a codimension-2 topological surface operator
\be
Q_e(\Sigma) = \int_{\Sigma_{D-2}} * J\,,
\ee
which generates an electric 1-form shift symmetry of the partially massless field. Similarly, due to~\eqref{Fbianchie} the magnetic field strength $*F_{\mu_1\cdots \mu_{D-2}\nu} = \frac{1}{2}\epsilon_{\mu_1\cdots \mu_{D}}F^{\mu_{D-1}\mu_D}_{\hphantom{\mu_{D-1}\mu_D}\,\nu}$ is also traceless and conserved on shell and can be used to construct a conserved $(D-2)$-form current $J_{D-2}=\ast J$ which can then be integrated as
\be
Q_m(\Sigma) = \int_{\Sigma_2} * J_{D-2}\,.
\ee
This charge generates a magnetic higher-form symmetry of the partially massless theory, which acts on the dual partially massless spin-2 by shifting it.
One can find charge and monopole solutions of the theory that act as the charged defects for the electric and magnetic higher form symmetries~\cite{Hinterbichler:2015nua}.

\subsection{Gauging fractonic symmetries}
The longitudinal mode of the partially massless graviton is precisely a galileon scalar~\cite{DeRham:2018axr}. When a massive graviton approaches the mass value \eqref{masstuning}, its degrees of freedom break up into those of a PM graviton and a scalar with the mass \eqref{Goldstonemasse}, so the Goldstone that arose in Section~\ref{symbreakingsubsection} is precisely the right mode that a PM graviton must combine with if it wants to get fully massive.

Recall that the field $\Phi$ transforms under the global symmetry as $\Phi\mapsto e^{i\alpha(x)}\Phi$, where $\alpha(x)$ must satisfy the reducibility condition~\eqref{eq:fractonreducibility}. We can remove this restriction and obtain a symmetry for  any $\alpha(x)$ by introducing a coupling to a partially massless gauge field $h_{\mu\nu}$,
\be
D^2_{\mu\nu}\Phi^2 \equiv  \nabla^2_{\mu\nu}\Phi^2 -i q\, M^\frac{6-D}{2} h_{\mu\nu}\Phi^2\,,
\ee
where $\nabla^2_{\mu\nu}\Phi^2$ is the covariant derivative given by~\eqref{eq:covariantD}, $M$ is the mass scale appearing in \eqref{eq:fractonaction}, and $q$ is a dimensionless coupling analogous to the electric charge. This is a gauge covariant version of the covariant derivative~\eqref{eq:covariantD}. Under the transformation
\be
\begin{aligned}
\Phi &\mapsto e^{iq\,\alpha(x)}\Phi\,,\\
h_{\mu\nu} &\mapsto h_{\mu\nu}+M^\frac{D-6}{2}\left( \nabla_\mu\nabla_\nu+H^2 g_{\mu\nu}\right)\alpha(x)\,,
\end{aligned}
\label{eq:gaugetransf2}
\ee
the gauge covariant derivative transforms covariantly 
\be D_{\mu\nu}^2 \mapsto e^{2iq\,\alpha(x)}D_{\mu\nu}^2,\ee
for {\it any} $\alpha(x)$. It is then straightforward to construct a gauge-invariant version of the action \eqref{eq:fractonaction} which couples $h_{\mu\nu}$ to $\Phi$\footnote{Although we do not utilize it here, it is also possible to couple the Wess--Zumino term~\eqref{eq:fractonicwzterm} to $h_{\mu\nu}$ in order to gauge the symmetry, though this only gauges the trace of the fracton's symmetry.
}
\be
\begin{aligned}
S = 
\int\rd^Dx \sqrt{-g}\Bigg[ 
&-\frac{1}{4}\left(F^{\lambda\mu\nu}F_{\lambda\mu\nu}-2F^{\lambda\mu}_{\ \ \ \mu} F_{\lambda\nu}^{\ \ \nu}\right)-\frac{1}{M^4}\Big(\lvert D^2_{\mu\nu}\Phi^2\rvert^2-\lvert D^2{}_\mu^{~\mu}\Phi^2\rvert^2\Big)\\
&+\left( \frac{1}{M^2}\Phi^*{}^2D^2{}_\mu^{~\mu}\Phi^2+{\rm c.c.}\right)+\mu^{D/2}\lvert\Phi\rvert^2-\frac{\lambda}{2}\lvert\Phi\rvert^4
\Bigg]\,,
\end{aligned}
\label{eq:gaugedfracton}
\ee
where we have also included the kinetic term~\eqref{PMkineticterme} for the PM spin-2 field. 

The terms linear in $h_{\mu\nu}$ in~\eqref{eq:gaugedfracton} are of the form $\sim h_{\mu\nu} J^{\mu\nu}$, where $J_{\mu\nu}$ is the doubly conserved tensor 
associated with the global fractonic symmetry mentioned at the end of Section~\ref{sec:fracton}. For this theory, it takes the form
\be 
J_{\mu\nu} \propto i \bar \Phi^2 \nabla^2_{\mu\nu}\Phi^2 - i g_{\mu\nu} \bar\Phi^2 {\nabla^2}_\rho^{\ \rho}\Phi^2 +{\rm c.c.} \ \ .\label{Jcurrenteforme}
\ee
Note that this coupling of the PM field to matter is not accounted for by the classification of and constraints on cubic couplings in~\cite{Joung:2012rv,Joung:2012hz,Sleight:2021iix} (which  imply certain conditions on the masses of scalars that can interact with a PM field) because those couplings assume a canonical kinetic term for all fields, which we do not have.  Similarly, the coupling to matter is not of the form studied in~\cite{Deser:2006zx} (which is actually a redundant cubic coupling that can be moved to higher order by a field redefinition, and thus does not mediate long range forces).

\subsection{Higgs phase}
\label{sec:Higgsphase}

We now want to consider the fate of this theory in the symmetry-breaking vacuum.
We expand~\eqref{eq:gaugedfracton} around the nontrivial solution~\eqref{eq:VEV}, again parameterizing the field as in~\eqref{eq:symmexpan}. At quadratic order in the fields, we find
\be
\begin{aligned}
S = \int\rd^Dx\sqrt{-g}\Bigg[&-\frac{1}{4}\left(F^{\lambda\mu\nu}F_{\lambda\mu\nu}-2F^{\lambda\mu}_{\ \ \ \mu} F_{\lambda\nu}^{\ \ \nu}\right)
-\frac{1}{2}\Delta m_h^2(h_{\mu\nu}^2-h^2)
-\frac{1}{2}Z^2_\phi\Big( (\nabla\phi)^2-DH^2\phi^2
\Big)\\
&+\kappa_{h\phi} h^{\mu\nu}\Big[\nabla_\mu\nabla_\nu\phi-g_{\mu\nu}(\square+(D-1)H^2)\phi
\big]
-\frac{1}{2}Z^2_\sigma\Big( (\nabla\sigma)^2+m_\sigma^2\sigma^2
\Big)
\Bigg]\,.
\end{aligned}
\label{eq:HiggsedPMs2}
\ee
The parameters $Z_\phi, Z_\sigma, m^2_\sigma$ are the same as in~\eqref{eq:znorm}--\eqref{eq:radialmass}, and the new parameters are
\begin{align}
\Delta m_h^2 &= \frac{2q^2\mu^D}{M^{D-2}\lambda^2}+\cdots\,,\\
\kappa_{h\phi} &= \frac{2\,q\,\mu^D}{M^\frac{D+2}{2}\lambda^2}+\cdots \,,
\end{align}
where the ellipses are terms suppressed by further powers of ${M^2\over H^2\lambda}$ or $\lambda$.

Under a gauge transformation~\eqref{eq:gaugetransf2}, $\sigma$ and $\phi$ transform as 
\be \delta \sigma=0\,,\ \qquad \ \delta \phi= q\alpha(x)\,.\label{Goldstoneshiftsgaugee}\ee
We see that the angular mode $\phi$ shifts, and serves as a St\"uckelberg field for the PM gauge invariance.  We can therefore completely fix the PM gauge symmetry by going to the analogue of unitary gauge where we set $\phi = 0$.  Fixing this gauge in the action~\eqref{eq:HiggsedPMs2} (which is permissible), it is then clear that the physical degrees of freedom are a massive spin-2 field $h_{\mu\nu}$ with Fierz--Pauli mass 
\be m_h^2 = (D-2)H^2+\Delta m_h^2,\ee
along with the radial mode $\sigma$ with the mass~\eqref{eq:radialmass}.  

We see that the graviton has become massive through a Higgs mechanism that starts from a PM field.  As in the ordinary Higgs mechanism, the angular mode $\phi$ mixes with the partially massless field into an irreducible representation, which is now massive. Note that all of $Z_\sigma^2, m_\sigma^2$ and $\Delta m_h^2$ are positive, so the theory is free of ghosts and tachyons, with $\Delta m_h^2$ giving the amount the massive graviton lies above the Higuchi bound.

Expanding to higher order in the fields gives interactions between the massive graviton and the radial mode.  To leading order in $H/ M$ these are, in unitary gauge,
\begin{align}
\nonumber
S_{\rm int} = \!\int\!\rd^Dx\sqrt{-g}\Bigg[&-\frac{q^2\,\mu^{D/2}}{M^{D-3}\lambda}\left(\hat\sigma+  {3M\lambda\over 8\mu^{D/2}}  \hat\sigma^2 +  {M^2\lambda^2\over 16\mu^{D}}  \hat\sigma^3 +  {M^3\lambda^3\over 256\mu^{3D/2}}  \hat\sigma^4   \right)\left( h_{\mu\nu}^2-h^2\right) \nn\\
&-\frac{\lambda^2M^3}{32\mu^{D/2}}\hat\sigma^3-\frac{\lambda^3M^4}{512\mu^D}\hat\sigma^4
-\frac{\lambda M}{4\mu^{D/2}}\hat\sigma(\nabla\hat\sigma)^2-\frac{3\lambda}{32M\mu^{D/2}}(\nabla\hat\sigma)^2\square\hat\sigma \nn\\
&-\frac{\lambda^2M^2}{32\mu^D}\hat\sigma^2(\nabla\hat\sigma)^2+\frac{\lambda^2}{64\mu^D}\hat\sigma\Big(\nabla_\mu\nabla_\nu\hat\sigma\nabla^\mu\hat\sigma\nabla^\nu\hat\sigma-(\nabla\hat\sigma)^2\square\hat\sigma
\Big) +\cdots\Bigg]\,.
\label{eq:fractonssbinteractions}
\end{align}
The self-interactions of $\hat\sigma$ are identical to~\eqref{eq:fractonssbinteractions}. The coupling to the graviton is through the Fierz--Pauli combination $h_{\mu\nu}^2-h^2$, which comes suppressed by $\sim (M^{D-2}\lambda)/q^2$.\footnote{If we had instead only included the coupling between $\Phi$ and $h_{\mu\nu}$ in the Wess--Zumino term \eqref{eq:fractonicwzterm}, the resulting theory would have no interactions between $\hat\sigma$ and $h_{\mu\nu}$, so that the resulting massive theory would effectively be a St\"uckelberging of a massive theory, with a decoupled scalar.}

\paragraph{Energy Scales} The mass of the radial Higgs mode is $m_\sigma^2 \sim \lambda M^2$, which is smaller than the cutoff scale of the fracton EFT $\sim M^2$ for $\lambda \ll 1$.  The mass of the graviton is $m_h^2\sim \frac{q^2}{\lambda^2} \left(\mu\over M\right)^{D-2}\mu^2$.  For this to be smaller than the cutoff scale we must have 
\be q^2 \ll \lambda^2 \left(\mu\over M\right)^D.\label{qcouplingregimee}\ee
Since both $\mu/M$ and $\lambda$ are small, this is a regime of weak coupling $q^2\ll 1$.

Much as in an ordinary superconductor, we would like to compare the mass of the Higgs mode to that of the massive gauge boson,
\be
\frac{m_h^2}{m_\sigma^2} \sim \frac{q^2}{\lambda^3}\left(\frac{\mu}{M}\right)^D\,.
\ee
There are two possible regimes that can both be realized while satisfying~\eqref{qcouplingregimee}:
\vspace{-6pt}
\begin{itemize}
\item {\bf Type I:} If $q^2 \gg \lambda^3\left(\frac{M}{\mu}\right)^D$, 
then the graviton is heavier than the radial mode.

\vspace{-6pt}
\item {\bf Type II:} If $q^2 \ll \lambda^3 \left(\frac{M}{\mu}\right)^D$,
then the  graviton is lighter than the radial mode.
\end{itemize}
\vspace{-4pt}
These are the PM analogues of the different types of superconductors.

\subsection{Relation to massive gravity}
In the Type II case, it is possible to integrate out the $\sigma$ field, yielding an EFT of just a massive graviton with the same strong coupling scale $\Lambda_s$ in \eqref{scsclaee} as the Goldstone theory.  Note that  the Type II regime can be parametrically realized consistently with~\eqref{qcouplingregimee} as well as all the constraints~\eqref{totalsinqee}.  For example, if we take $q\sim \lambda^{t_3}$, then~\eqref{qcouplingregimee} and the Type II condition become the inequalities
\be 
2 t_3 > 2 +  Dt_2\, ,\ \qquad \  2 t_3 < 3 - Dt_2 .
\ee
When these are combined with~\eqref{simutineeqee}, there exists for all $D\geq 3$ a region in the space of positive $t_1,t_2,t_3$ where all constraints are satisfied.  One can take $t_3\rightarrow \infty$ within this region, which sends $q\rightarrow 0$ and recovers the un-gauged theory.

While the theory obtained by integrating out $\sigma$ is nonlinear and propagates the correct number of degrees of freedom of a massive graviton below $\Lambda_s$, it is {\it not} the effective field theory of a massive graviton with the Einstein--Hilbert kinetic structure of GR (and so in particular is not the dRGT theory~\cite{deRham:2010kj}). Instead it has only the linear Fierz--Pauli kinetic term, and so it is an example of ``pseudo-linear" massive gravity~\cite{Folkerts:2011ev,Hinterbichler:2013eza,Bonifacio:2018van} on dS.

Since the massive gravity EFT arises from integrating out the $\sigma$ field, it will have a cutoff $\sim m_\sigma$ which appears suppressing higher derivative interactions. If we restore the longitudinal mode $\phi$ via the St\"uckelberg trick it will be described at high energies by the Goldstone EFT obtained from~\eqref{eq:fractonssbinteractions} by integrating out the radial mode. The strong coupling scale of this EFT is $\Lambda_s$, and the strong coupling scale of the graviton EFT should be the same.  As we noted above, this strong coupling scale can be made parametrically larger than the graviton mass, and parametrically smaller than the UV scale $M$. 
This bottom up perspective provides an interesting take on the construction. We see that this should give an example where the strong coupling scale of a theory of an isolated massive spin-2 field can be raised by integrating in additional lower spin matter fields at tree level, something which cannot be done in flat space \cite{Bonifacio:2019mgk}. The novelty of this situation is that the theory in the ultraviolet is that of a partially massless spin-2 field with non-standard fractonic matter.

\subsection{Quasi-topological field theory description}

It is interesting to consider a more symmetry-centric viewpoint on the PM Higgs phase.  As reviewed at the end of Section~\ref{PMsymmsection}, the free partially massless spin-2 field itself has 
both electric and magnetic higher-form symmetries \cite{Hinterbichler:2015nua}. The electric symmetry is a $1$-form symmetry, with corresponding conserved current $F_{\mu\nu\lambda}$,~\eqref{eq:PMmaxwell}, while the magnetic symmetry is a 
$(D-3)$-form symmetry, with a corresponding conserved current  $J_{\mu_1\cdots\mu_{D-2}\,\nu} = \frac{1}{2}\epsilon_{\mu_1\cdots\mu_D}F^{\mu_{D-1}\mu_{D}}_{~~~~~~~~~~\nu} $. 
This current is traceless and
has the index symmetries of the Young diagram
\be
J_{\mu_1\cdots\mu_{D-2}\,\nu} \in ~D-2\,
\Bigg\{\,
\hspace{-3pt}
\raisebox{11.5pt}{
\gyoung(~;~,|2\vdts)}
\label{eq:Jsymmetry} \,\ \ . 
\ee
On shell, both currents $F$ and $J$ are traceless and conserved in all their indices.

In addition to this, the matter sector has the U$(1)$ $0$-form symmetry \eqref{eq:fractonsymm}, which has a corresponding current \eqref{Jcurrenteforme} that is a two-index symmetric tensor $J_{\mu\nu}$ found from reading off the terms linear in $h_{\mu\nu}$ in the action \eqref{eq:gaugedfracton}.  It obeys the double conservation condition $(\nabla^\mu\nabla^\nu+H^2g^{\mu\nu}) J_{\mu\nu} = 0$. 

In the free PM theory, the higher-form symmetries are nonlinearly realized, with the partially massless graviton $h_{\mu\nu}$ serving as the Goldstone mode for these symmetries.\footnote{This is analogous to  familiar examples in flat space, like that of electromagnetism~\cite{Rosenstein:1990py,Kovner:1992pu,Gaiotto:2014kfa}, and gravity~\cite{Benedetti:2021lxj,Hinterbichler:2022agn,Benedetti:2023ipt,Hinterbichler:2024cxn}.}
Coupling the partially massless field directly to matter, as we do in~\eqref{eq:gaugedfracton}, explicitly breaks the electric symmetry, but the magnetic symmetry persists. In addition, this coupling gauges the fractonic dipole symmetry.
In the ordinary ``Coulomb" phase, where the graviton is partially massless, the PM magnetic symmetry is non-linearly realized.  On the other hand, in the Higgs phase, the PM magnetic symmetry is linearly realized (much as in the superconducting phase of electromagnetism~\cite{Rosenstein:1990py,Kovner:1992pu}). 

The natural observables of the fracton-PM system are correlation functions of the magnetic current \eqref{eq:Jsymmetry}. We can characterize these by coupling the system to a background gauge field $A_{\mu_1\cdots\mu_{D-2}\,\nu}$.
This gauge field has the same symmetry type as $J$ in~\eqref{eq:Jsymmetry}, and can be taken to be traceless since only the traceless part of the magnetic current is nontrivial on-shell.  This background source couples via $\sim \epsilon^{\mu_1\cdots\mu_D}A_{\mu_1\cdots\mu_{D-2}\,\nu} F_{\mu_{D-1}\mu_D}^{\hphantom{\mu_{D-1}\mu_D}\nu}$ and it transforms as 
\be
\delta_\xi A_{\mu_1\cdots\mu_{D-2}\,\nu} =  {\cal P}_T\nabla_{[\mu_1}\xi_{\mu_2\cdots\mu_{D-2}]\,\nu}\,,
\label{eq:agauge}
\ee
where $\xi_{\mu_1\cdots\mu_{D-3}\,\nu}$ is antisymmetric in its first $D-3$ indices, and vanishes if we antisymmetrize all of its indices, and the projector ${\cal P}_T$ projects onto the fully traceless part.

Now, we can imagine integrating out 
the dynamical degrees of freedom to obtain an effective action for the background fields which serves as a generating functional for correlation functions.
In the ordinary phase, the fracton lacks a standard kinetic term, and so the resulting effective action would be a bit hard to characterize. Additionally, the presence of a PM field will lead to it being nonlocal at distances of order the de Sitter length.
 However, in the Higgs phase, all the degrees of freedom are massive, and so the resulting effective theory will be local at energies below all the mass scales. In fact, since we have integrated out all the propagating degrees of freedom, it will just be a theory of ground states, with no remaining propagating degrees of freedom.  
 
We can obtain this description from the Landau--Ginzburg-like description discussed in Section~\ref{sec:Higgsphase} by considering the limit where the masses of all the particles in the Higgs phase are scaled to infinity. In this limit, the equations of motion set $h_{\mu\nu} = 0$ in unitary gauge. We can reproduce this constraint with the effective action
\be
S = \int\rd^Dx\sqrt{-g}\bigg[\Big( q\,h_{\mu\nu}+\nabla^2_{\mu\nu}\phi\Big) \lambda^{\mu\nu} - \epsilon^{\mu_1\cdots\mu_D}A_{\mu_1\cdots\mu_{D-2}\,\nu} F_{\mu_{D-1}\mu_D}^{\hphantom{\mu_{D-1}\mu_D}\nu}
\bigg]\,,
\label{eq:fractonLG}
\ee
where $\lambda^{\mu\nu}$ is a Lagrange multiplier, we have reintroduced $\phi$ as a St\"uckelberg field.
(Note that the fields here are normalized differently from~\eqref{eq:fractonssbinteractions},
and we have pulled out the dimensionless charge $q$ explicitly.) In addition, since the magnetic symmetry is present in the Higgs phase, we have included the coupling to the background gauge $A_{\mu_1\cdots\mu_{D-2}\,\nu}$ corresponding to the magnetic symmetry. 
The fields $h_{\mu\nu}$, $\phi$ are inert under the background gauge transformation \eqref{eq:agauge}.

The action~\eqref{eq:fractonLG} describes the long-distance dynamics of the Higgs phase. In order to imbue this system with richer phenomenology, it is useful to parallel the TQFT description of superconductivity~\cite{Maldacena:2001ss,Hansson:2004wca,Banks:2010zn,Gukov:2013zka,Thorngren:2023ple} and
contemplate the situation where we have some additional structure that allows us to notice that the matter sector has a U$(1)$ symmetry of the form~\eqref{eq:fractonsymm}. This symmetry of $\Phi$ is gauged by the coupling to $h_{\mu\nu}$, but we could consider e.g. introducing additional fields charged under this symmetry which do not couple to $h_{\mu\nu}$, or have $\Phi$ carry a multiple of the fundamental charge of the system. In this case, there will be an additional matter conserved current $J_{\mu\nu}$, which satisfies $(\nabla^\mu\nabla^\nu+H^2g^{\mu\nu}) J_{\mu\nu} = 0$, and it is natural to introduce a gauge field $B_{\mu\nu}$ to source this current.  It couples via $\sim B_{\mu\nu}J^{\mu\nu}$ and it transforms as 
\be
\label{eq:bgauge}
\delta_\Lambda B_{\mu\nu} = \left(\nabla_{\mu}\nabla_{\nu}+H^2g_{\mu\nu}\right)\Lambda\,,
\ee
with a scalar gauge parameter $\Lambda(x)$.

If we assume that $\phi$ carries charge $p$ under this additional matter U$(1)$, then we can introduce the gauge field $B_{\mu\nu}$  into~\eqref{eq:fractonLG} as
\be
S = \int\rd^Dx\sqrt{-g}\bigg[\Big( q\,h_{\mu\nu}+p\,B_{\mu\nu}+\nabla^2_{\mu\nu}\phi\Big) \lambda^{\mu\nu} - \epsilon^{\mu_1\cdots\mu_D}A_{\mu_1\cdots\mu_{D-2}\,\nu} F_{\mu_{D-1}\mu_D}^{\hphantom{\mu_{D-1}\mu_D}\nu}
\bigg]\,.
\ee
This is invariant under \eqref{eq:bgauge} where the $\phi$ field also transforms as $\delta\phi = -p\,\Lambda(x)$.  This non-trivial transformation of $\phi$ mirrors the transformation \eqref{Goldstoneshiftsgaugee} of the full theory. 
We can now formally integrate out $h_{\mu\nu}$ using the constraint enforced by $\phi$ to obtain a BF theory~\cite{Horowitz:1989ng}:\footnote{This can be subtle if $p/q$ is fractional, due to an additional discrete $1$-form symmetry. This will not be important for our purposes, but could be interesting to explore. See~\cite{Thorngren:2023ple} for a discussion in the ordinary superconductor case.}
\be
S = \frac{p}{q}\int\rd^Dx\sqrt{-g}\epsilon^{\mu_1\cdots \mu_D}A_{\mu_1\cdots\mu_{D-2}\,\nu} \nabla_{\mu_{D-1}}B_{\mu_{D}}^{~~~\nu}\,.
\label{eq:SBFB}
\ee
For our purposes, it is convenient to integrate by parts (equivalently add a suitable boundary term) to put the action in the form
\be
S = -\frac{p}{q}\int \rd^Dx\sqrt{-g}\, \epsilon^{\mu_1\cdots \mu_D}B_{\mu_1}^{~~\lambda}\, F^{(m)}_{\mu_2\cdots\mu_D\,\lambda}\,,
\label{eq:tqft}
\ee
where $F^{(m)}_{\mu_1\cdots\mu_{D-1}\nu}$ is the gauge invariant field strength associated to $A_{\mu_1\cdots\mu_{D-2}\,\nu}$~\cite{Hinterbichler:2024cxn},
\be
F^{(m)}_{\mu_1\cdots\mu_{D-1}\nu} \equiv  {\cal P}_T\,\nabla_{[\mu_1}A_{\mu_2\cdots\mu_{D-1}]\,\nu}\,.
\ee

The action~\eqref{eq:tqft} is a partially massless BF theory.  It is gauge invariant under the transformations~\eqref{eq:agauge}, \eqref{eq:bgauge}. Though we have derived this Lagrangian from the Landau--Ginzburg description, we could have written this down directly from the symmetries of the problem as the lowest-order gauge-invariant effective action for the conserved currents.
A small twist compared to the usual $p$-form BF theories is that in this case it is necessary to raise an index in order to write the action~\eqref{eq:tqft}. This means that this theory requires a metric to define, unlike the usual BF theories, and so it is not strictly topological in the way that those theories are.  Nevertheless, it has no local propagating degrees of freedom, and as such we call it quasi-topological (similar constructions in the flat space context can be found in \cite{Bertolini:2023sqa,Huang:2023zhp,Bertolini:2024yur,Rovere:2024nwc,Bertolini:2025qcy,Bertolini:2025jul}).

\paragraph{Edge Modes:}
Since there are no propagating degrees of freedom, BF actions are locally trivial in an empty bulk.  But they can describe non-trivial effects on the boundary of manifolds with boundary~\cite{Maldacena:2001ss,Kravec:2013pua,Fliss:2023uiv}. In de Sitter space, where~\eqref{eq:tqft} is defined, there is a natural boundary located at future infinity. We would like to understand the consequences of~\eqref{eq:tqft} on this boundary. 

Under large gauge transformations (i.e. those whose gauge parameter does not vanish on the boundary), the action~\eqref{eq:tqft} is invariant under~\eqref{eq:agauge}, but not~\eqref{eq:bgauge}.\footnote{This is a choice we made by writing the action as~\eqref{eq:tqft}. If we add a boundary term to write the action in terms of  the field strength of $B$, it will be exactly invariant under~\eqref{eq:bgauge} and would transform by a boundary term under~\eqref{eq:agauge}. This inability to have {\it both} gauge invariances simultaneously signals a mixed 't Hooft anomaly between the matter U$(1)$ symmetry and the magnetic PM symmetry: the failure of gauge invariance can be moved around but cannot be eliminated entirely.} Instead it transforms by
the boundary term
\be
\begin{aligned}
\delta_\Lambda S &= \frac{p}{q}\int \rd^Dx\sqrt{-g}\, \nabla_{\mu_1} \left(\epsilon^{\mu_1\cdots \mu_D} \nabla^\lambda\Lambda\, F^{(m)}_{\mu_2\cdots\mu_D\,\lambda}\right)\,\\
&= \frac{p}{q}\int \rd^dx\sqrt{\gamma}\,\epsilon^{i_1\cdots i_{D-1}}\,\Lambda  \nabla^j\nabla_{i_1} A_{i_2\cdots i_{D-1}\, j}\,,
\label{eq:anomvar}
\end{aligned}
\ee
where $i,j,\ldots$ are indices of the $d=D-1$ dimensional boundary and we have imposed as boundary conditions that the non-tangential components of both the gauge fields and the gauge parameters vanish on the boundary.

The anomalous variation~\eqref{eq:anomvar} signals the presence of an edge mode that cancels the variation via anomaly inflow. The edge mode theory must be a boundary theory that has the same conserved current content and anomaly structure as the bulk theory.  The question of how to build a theory with a prescribed set of currents and anomalies between them is precisely the question addressed in~\cite{Hinterbichler:2024cxn}, and we can use the methods described there to find the boundary theory.  The boundary theory in this case turns out to be a scalar with a higher-derivative kinetic term, of the type studied in~\cite{Brust:2016gjy}.  It is easiest to describe the theory in the flat slicing of de Sitter so that the boundary metric is flat. Then, the gapless boundary theory has the action
\be
S = -\int\rd^d x\, \frac{1}{2}\phi\, \square^2\phi\,.
\ee
As discussed in~\cite{Hinterbichler:2024cxn} (see also~\cite{Nicolis:2010se}), this theory possesses an (electric) conserved current
\be
J_{ij} = \partial_{(i}\partial_{j)_T}\phi\,,
\ee
which satisfies the double conservation condition $\partial^i\partial^j J_{ij} = 0$, as well as a (magnetic) higher-form conserved current\footnote{In the free theory the magnetic and electric currents are hodge duals, but this will not be true with interactions.}
\be
K_{i_1\cdots i_{d-1}}^{\hphantom{i_1\cdots i_{d-1}}\,j} = \epsilon_{i_1\cdots i_{d}} \partial^{(i_d}\partial^{j)_T}\phi\,.
\ee
We can then introduce background fields for each of these currents, which can be thought of as the spatial components of $B_{\mu\nu}$ and $A_{\mu_1\cdots\mu_{D-2}\,\nu}$.\footnote{Though $B_{\mu\nu}$ is not traceless, its trace projects out of~\eqref{eq:SBFB} so only the traceless part is relevant for the anomaly.}  There is a mixed anomaly between conservation of the electric and magnetic currents~\cite{Hinterbichler:2024cxn}, so it is impossible to construct an action that is gauge invariant under transformations of both background fields. Under the background field gauge transformations, $\phi$ transforms as $\delta_\Lambda\phi = \Lambda$, and $\delta_\xi\phi = 0$.
An action that is invariant under the magnetic gauge transformation is
\be
S = -\int\rd^d x\left[\frac{1}{2}(\partial_{(i}\partial_{j)_T}\phi-B_{ij})^2+\frac{p}{q}\epsilon^{i\, i_2\cdots i_d}A_{i_2\cdots i_d}^{\hphantom{i_2\cdots i_d}\,j}\partial_{(i}\partial_{j)_T}\phi\right]\,.
\ee
Under the electric gauge transformation, it is not invariant, but rather transforms as
\be
\delta_\Lambda S = - \frac{p}{q}\int\rd^d x\left(\epsilon^{i\, i_2\cdots i_d}\Lambda \,\partial_i\partial_jA_{i_2\cdots i_d}^{\hphantom{i_2\cdots i_d}\,j}\right)\,,
\ee
where we have integrated by parts. Comparing with~\eqref{eq:anomvar} we see that this anomalous variation exactly cancels so that the combination of bulk plus boundary actions is gauge invariant. Thus we see that---much like an ordinary superconductor---the Higgsed partially massless graviton phase supports a gapless edge mode localized on the boundary.

\paragraph{Persistent Current:}
One of the characteristic features of a superconducting phase is the presence of persistent, dissipationless currents in the presence of a background chemical potential.\footnote{Also important is the Meissner effect---the vanishing static magnetic susceptibility of the system. This follows fairly straightforwardly from the effective description. The leading term quadratic in $A$ is $F_{(m)}{}^2$. The magnetic response given by differentiating this twice with respect to $A$ vanishes like $p^2$ at small momentum.} We can understand the presence of such persistent edge currents in the partially massless superconductor from~\eqref{eq:tqft}, following~\cite{PhysRevB.104.205132,Thorngren:2023ple}. Differentiating~\eqref{eq:tqft} with respect to $B_{\mu\nu}$, we obtain the expectation value of $J_{\mu\nu}$ in the presence of background gauge fields:
\be
\langle J_{\alpha\beta}\rangle = - \frac{p}{q} \epsilon_{\alpha\mu_2\cdots \mu_D}\, F_{(m)}^{\mu_2\cdots\mu_D}{}_\beta\,.
\ee
Defining the dual source $E_{\alpha\beta}\equiv \epsilon_{\alpha\mu_2\cdots \mu_D}\, F_{(m)}^{\mu_2\cdots\mu_D}{}_\beta $, we can write
\be
\langle J_{ij}\rangle  = - \frac{p}{q} E_{ij}\,.
\ee
This implies that in the presence of a background partially massless electric field source, there is an induced current of fractonic charges. Similarly, turning on a background $B_{\mu\nu}$ field induces a magnetic current.

\section{Higher-spin generalization}
\label{sec:HS}

Thus far we have considered the Higgsing of a partially massless spin-2, which is the simplest example of a partially massless field, but there exist partially massless fields of every spin $\geq 2$. In fact the structure becomes richer as we go up in spin:
a spin-$s$ field has $s-1$ partially massless points in addition to the massless point, together labeled by an integer $t$ (called the depth) which runs between $0\leq t\leq s-1$ with $t=s-1$ the massless case.   The spin $s$ depth $t$ field occurs at the mass value
\be
 m_{s,t}^2 = (s-t-1)(s+t+D-4)H^2\,,
\ee
where the helicities $0,1,\ldots,t $ are projected out by the presence of a $(s-t)$-derivative gauge invariance with a $t$-index gauge parameter. 

The partially massless spin-2 is a member of the family of maximal depth ($t=0$) partially massless theories, and shares many features with its higher-spin cousins. It is therefore natural to consider generalizing the constructions above to spin-$s$ theories of maximal depth, which can be expected to give relativistic dS versions of higher multipole fractonic theories \cite{Gromov:2018nbv,Stahl:2021sgi}.

Here we sketch the construction. At the mass values $ m_{s,0}^2 = (s-1)(s+D-4)H^2$, the spin-$s$ field $\ell_{\mu_1\cdots\mu_s}$ develops a scalar gauge invariance
\be
\delta \ell_{\mu_1\cdots\mu_s} = {\cal D}^s_{\mu_1\cdots\mu_s} \alpha\,,
\ee
where we have defined the differential operator:
\be
{\cal D}^s_{\mu_1\cdots\mu_s}  \equiv \begin{cases}
{\cal Y}_{[s]}\left(\prod_{n=1}^{\frac{s}{2}}\left[\nabla_{\mu_n}\nabla_{\mu_{n+\frac{s}{2}}}+(2n-1)^2H^2g_{\mu_n\mu_{n+\frac{s}{2}}}\right]\right) &{\rm for}~s~{\rm even} ,\\[8pt]
{\cal Y}_{[s]}\left(\prod_{n=1}^{\frac{s-1}{2}}\left[\nabla_{\mu_n}\nabla_{\mu_{n+\frac{s-1}{2}}}+(2n)^2H^2g_{\mu_n\mu_{n+\frac{s-1}{2}}}\right]\right)\nabla_{\mu_{s}} &{\rm for}~s~{\rm odd},
\end{cases}
\ee
where ${\cal Y}_{[s]}$ is a projector onto the totally symmetric index combination. This scalar gauge invariance removes the helicity-$0$ polarization so that the field propagates  $(D-3+2s)(D-4+s)!/[s!(D-3)!] - 1$ degrees of freedom---one fewer than a massive spin-$s$ particle. We can restore this missing degree of freedom by coupling to a higher-multipole fracton theory. We consider the covariant derivative
\be
\nabla^s_{\mu_1\cdots \mu_s}\Phi^s \equiv \Phi^s {\cal D}^s_{\mu_1\cdots\mu_s}\log\left(\frac{\Phi}{f}\right)\,.
\ee
Under the transformation $\Phi\mapsto e^{iq\,\alpha(x)}\Phi$, this transforms as
\be
\nabla^s_{\mu_1\cdots \mu_s}\Phi^s \mapsto e^{i sq\,\alpha(x)}\nabla^s_{\mu_1\cdots \mu_s}\Phi^s+ e^{i sq\,\alpha(x)}\Phi^s {\cal D}^s_{\mu_1\cdots\mu_s}\alpha(x)\,.
\label{eq:spinsglobal}
\ee
So, we see that this theory transforms covariantly under a global phase rotation if  ${\cal D}^s_{\mu_1\cdots\mu_s}\alpha(x) = 0$.\footnote{The solutions of this equation can be written in terms of traceless symmetric polynomials in embedding space.} We can then construct an invariant action by contracting together powers of the covariant derivative $\nabla^s_{\mu_1\cdots \mu_s}\Phi^s$. Expanding around a symmetry breaking minimum, the angular mode will have a fixed mass $m_\phi^2 = -(s-1)(s+D-2)H^2$, which is the level $k=s-1$ shift-symmetric scalar mass value studied in~\cite{Bros:2010wa,Epstein:2014jaa,Bonifacio:2018zex}. In order for this field to have a standard kinetic term, the precise quadratic contractions of the covariant derivative $\nabla^s_{\mu_1\cdots \mu_s}\Phi^s$ should be tuned in the same way as the mass terms in the Singh--Hagen Lagrangian~\cite{Singh:1974qz} for a massive spin $s$, extended to dS \cite{Hallowell:2005np}.

We can gauge the symmetry~\eqref{eq:spinsglobal} by coupling to the partially massless field through
\be
D^s_{\mu_1\cdots \mu_s}\Phi^2 \equiv \nabla^s_{\mu_1\cdots \mu_s}\Phi^2 -i q\, \ell_{\mu_1\cdots\mu_s} \,,
\ee
which now transforms covariantly for general $\alpha(x)$. Following the same procedure as in Section~\ref{sec:Higgsphase}, we construct a gauge-invariant theory of the partially massless field and the fracton which has a symmetry breaking potential and finite density ground state. Expanding around this vacuum, the angular Goldstone will mix with the partially massless field, Higgsing it into a massive spin-$s$ field.

We can also consider the topological description of the far infrared of the theory. A spin-$s$ depth-$0$ partially massless field $\ell_{\mu_1\cdots\mu_s}$ possesses a gauge-invariant curvature which is given by
$
F^{(s)}_{\nu\mu_1\cdots\mu_s} = \nabla_{\nu}\ell_{\mu_1\cdots\mu_s}-\nabla_{\mu_1}\ell_{\nu\cdots\mu_s}\,,
$
that is both conserved and closed on-shell. Thus, the curvature and its dual can be thought of as the conserved currents of electric and magnetic higher-form symmetries. In the Higgs phase, the electric symmetry is explicitly broken, but the magnetic symmetry is preserved. In addition to this, there is a $0$-form fractonic symmetry of the matter sector with a corresponding symmetric  conserved current $J_{\mu_1\cdots\mu_s}$, which satisfies  ${\cal D}_s^{\mu_1\cdots\mu_s}J_{\mu_1\cdots\mu_s} = 0$. If we integrate out all the degrees of freedom in the massive phase, the resulting effective description is a BF theory of the form
\be
S \propto \int\rd^Dx\sqrt{-g}\, \epsilon^{\mu_1\cdots\mu_D} B_{\mu_1}^{~~\nu_1\cdots\nu_{s-1}}\partial_{\mu_2}A_{\mu_3\cdots\mu_{D}\,\nu_1\cdots\nu_{s-1}}\,,
\label{eq:hsBF}
\ee
where $B_{\mu_1\cdots \mu_s}$ is a symmetric gauge field, which sources the matter current, and $A_{\mu_1\cdots\mu_{D-2}\,\nu_1\cdots\nu_{s-1}}$ is a traceless gauge field with the symmetries of the Young diagram
\be
J_{\mu_1\cdots\mu_{D-2}\,\nu} \in ~D-2\,
\Bigg\{\,
\hspace{-3pt}
\raisebox{11.5pt}{
\gyoung(_4{s},|2\vdts)}\,.
\label{eq:spinsJsymmetry}
\ee
 This latter current sources the magnetic current $\epsilon^{\alpha_1\cdots\alpha_{D-2}\nu\mu_1}F^{(s)}_{\nu\mu_1\cdots\mu_s}$
The action~\eqref{eq:hsBF} is invariant under the background gauge transformations
\begin{align}
\delta B_{\mu_1\cdots\mu_s} &= \nabla_{(\mu_1}\cdots\nabla_{\mu_s)_T}\Lambda(x)\\
\delta A_{\mu_1\cdots\mu_{D-1}\,\nu_1\cdots\nu_{s-1}} &= {\cal Y}^T_{[s,1,\cdots, 1]}\nabla_{\mu_1}\xi_{\mu_2\cdots\mu_{D-1}\,\nu_1\cdots\nu_{s-1}}\,,
\end{align}
where $\xi$ is a gauge parameter with one fewer row than $A$ and ${\cal Y}^T_{[s,1,\cdots, 1]}$ is a Young projector onto the tensor with the same index symmetries as $A$. 

On a manifold with boundary, like de Sitter space with its future spacelike boundary, the action~\eqref{eq:hsBF} is not invariant under large gauge transformations.  Again there is a mixed 't Hooft anomaly and we can  make a choice of which gauge invariance is preserved.  In this presentation, it is the matter gauge invariance parameterized by $\Lambda$ that is anomalous. The anomalous variation can be canceled by a gapless edge mode theory of the form
\be
S \propto \int\rd^dx\, \phi\, \square^s\phi\,.
\ee
One can similarly derive persistent multipole currents emblematic of the superconducting phase.

\section{Conclusions}
\label{sec:conclusions}

We have described a Higgs mechanism for the partially massless graviton on de Sitter, and its generalization to maximal depth higher spin PM fields. 
Given the similarities between depth $t=0$ partially massless fields and electromagnetism, this deserves to be called a partially massless superconducting phase.  Indeed, we have seen that this Higgs phase shares many phenomenological similarities with superconductors, including gapless edge modes and persistent currents.  Aside from serving as the first (to our knowledge) example of a covariant Higgsing of a relativistic field of spin $\geq 2$, and providing a look into what the phase structure of a gravitational theory could look like, the analysis raises a number of interesting questions to pursue.

We have described the superconducting phase transition via a Landau--Ginzburg description which is itself an EFT. It is natural to wonder if there is a more fundamental description of the condensation of the fractons themselves in the spirit of BCS theory. Relatedly, it would be interesting to understand if the partially massless fields can also arise as Goldstone modes for higher-form symmetry breaking related to the condensation of extended objects along the lines of~\cite{Polyakov:1980ca,Iqbal:2021rkn}. Since the magnetic symmetry is preserved in the Higgs phase, it would be interesting to directly construct the magnetically charged vortex solutions and study their properties.

Another viewpoint on the construction described here is that it is a mechanism to consistently quantize the scalar exceptional series representations~\cite{Bros:2010wa,Epstein:2014jaa,Bonifacio:2018zex}. These theories possess shift symmetries, which lead to zero modes that naively break de Sitter symmetry, and so their quantization is subtle. One approach is to treat the shift symmetries as gauge invariances~\cite{Epstein:2014jaa}. Here we are essentially considering a different possibility, that one should quantize the theories by embedding them inside a massive spin-$s$ representation. This is philosophically similar to the approach taken in~\cite{Anninos:2023lin} in $1+1$ dimensions. It would be nice to find and elucidate a UV representation of this construction. In particular, since the fractonic field is not well defined around $\Phi = 0$, it is not clear that it has a Fock space-like construction, and so it is somewhat obscure what its fundamental formulation looks like.

Our analysis was limited to maximal-depth partially massless fields. It is natural to wonder if a similar construction can be employed to Higgs the other depths of partially massless fields. The most interesting example would be the massless graviton. Following the logic described in Section~\ref{sec:PMHiggs}, we should expect that the matter theories that would be able to Higgs these fields would have global symmetries that are the same as the reducibility parameters of  the gauge transformations, which in this case would be the Killing vectors of dS. This suggests that the linearized theories can be Higgsed by some version of a ``vector fracton." Theories of this type have been studied (see, e.g.,~\cite{Pretko:2018jbi}), though it is difficult to make them covariant. A further complication is that any such Higgs mechanism would have to give the fields sufficient mass to push them above the Higuchi bound in order to be stable in de Sitter space.\footnote{In AdS, there is no such problem and Higgsing can happen at one-loop via coupling to a CFT through the mechanism of~\cite{Porrati:2001db,Porrati:2001gx,Porrati:2003sa}.  In flat space there is no way to add lower-spin particles to a theory of massive gravity at tree level in order to raise its strong coupling scale~\cite{Bonifacio:2019mgk}, which precludes the kind of Higgsing we are discussing here.} Most ambitiously, we could imagine trying to Higgs fully non-linear Einstein gravity, and study its phase structure. Given the nonlinear diffeomorphism gauge invariance of such a theory, it suggests that we would need a matter sector with a global diffeomorphism symmetry---a kind of fluid~\cite{Anderson:1963pc}. One can then view the challenges of Higgsing Einstein gravity as being related to the difficulty of finding a fluid phase that preserves relativistic symmetries (though see~\cite{Esposito:2020wsn} for an inspiring construction).

From the more phenomenological side, we have seen that massive spin-$s$ fields, if they arise from Higgsing an underlying partially massless field, can retain some memory of their origins through the presence of edge modes. In the inflationary context, such degrees of freedom would be localized at the end of inflation, on the reheating surface. The propagating degrees of freedom of a massive spin-$s$ field can leave interesting signatures in cosmological correlations~\cite{Arkani-Hamed:2015bza,Arkani-Hamed:2018kmz,Baumann:2019oyu}. It is natural to wonder if the edge modes studied here can leave similarly interesting imprints to be decoded in cosmological measurements. Since they are gapless, edge modes might be visible even if the massive fields with which they are associated are too heavy to be seen. More generally, it will be interesting to further study the properties of the partially massless BF theories~\eqref{eq:hsBF} in de Sitter space and to understand any connections with the edge modes seen in the static patch~\cite{Anninos:2020hfj,Law:2020cpj,Ball:2024hqe,Law:2025ktz}.

Finally, we have focused on physics in de Sitter space, but one can equally well euclideanize everything and consider the construction on the sphere. One lesson from this analysis is that relativistic fracton theories appear to be better behaved on the sphere as compared to flat space (where we saw that the Goldstone theory becomes infinitely strongly coupled). It is therefore natural to wonder whether either the relativistic fracton theory, or possibly an analogue of a partially massless superconductor, could be engineered on a spherical sample in a laboratory setting.

Here we have seen a first glimpse of the phase structure of a quasi-gravitational theory. It both bears natural resemblances to electromagnetism, and has new intriguing features. We hope that it can serve as a guidepost to progress on similar questions in the true gravitational context.

\vskip7pt
\noindent
\textbf{Acknowledgements:} We thank Dionysios Anninos, Tarek Anous, Jackson Fliss, Diego Hofman, and Albert Law for helpful discussions.  KH acknowledges support from DOE award
DE-SC0009946.  AJ is supported by DOE award DE-SC0025323 and by the Kavli Institute for Cosmological Physics at the University of Chicago.

\vspace{-4pt}
\linespread{.8925}
\addcontentsline{toc}{section}{References}
\bibliographystyle{utphys}
{\small
\bibliography{PMhiggs_arxiv}
}

\end{document}